\documentclass{article}
\usepackage{ijcai21}
\usepackage{soliditystyler}
\usepackage{tracestyler}

\usepackage{times}
\usepackage{soul}
\usepackage{url}
\usepackage[hidelinks]{hyperref}
\usepackage[utf8]{inputenc}
\usepackage[small]{caption}
\usepackage{graphicx}
\usepackage{amsmath}
\usepackage{amsthm}
\usepackage{booktabs}
\usepackage{algorithm}
\usepackage{algorithmic}
\usepackage{color}
\title{Dynamic Vulnerability Detection on Smart Contracts Using Machine Learning}

\author{
Mojtaba Eshghie$^1$\and 
Cyrille Artho$^1$\and
Dilian Gurov$^1$
\affiliations
$^1$
School of Electrical Engineering and Computer Science, KTH Royal Institute of Technology\\
\emails
\{eshghie, artho, dilian\}@kth.se
}

\begin{document}

\maketitle

\begin{abstract}
In this work we propose Dynamit, a monitoring framework to detect reentrancy vulnerabilities in Ethereum smart contracts.
The novelty of our framework is that it relies only on transaction metadata and balance data from the blockchain system;
our approach requires no domain knowledge, code
instrumentation, or special execution environment.
Dynamit extracts features from transaction data and uses a machine learning model to classify transactions as benign or harmful.
Therefore, not only can we find the contracts that are vulnerable to reentrancy attacks, but we also get an execution trace that reproduces the attack.
Using a random forest classifier,
our model achieves more than 90 percent accuracy on 105 transactions, showing
the potential of our technique.
\end{abstract}

\section{Introduction}
A blockchain is a distributed ledger that manages assets between users.
A \emph{smart contract} 
encodes rules to handle the transfer of these
assets. The transfers happen within transactions that are
stored on the blockchain and are persistent. Therefore, smart contracts can
implement a wide range of use cases, including financial and governance
applications~\cite{luuMakingSmartContracts2016a}. For instance,
a contract could act like an autonomous agreement between multiple
parties to transfer assets to desired accounts when
particular conditions are met.

One of the most popular blockchain platforms that supports smart
contracts is Ethereum. As of January 2021, the market capitalization of
Ethereum is about 130 billion dollars~\cite{CryptocurrencyPricesCharts}. Smart
contracts on Ethereum allow users and other
contracts to interact with them by making
calls to functions in the contracts. 
Smart contracts in Ethereum are commonly written in Solidity, a language influenced by JavaScript. 
To perform
the operations on a smart contract, an execution fee is needed that is
called \emph{gas}. Gas fees are paid in Ethereum's native currency,
\emph{Ether} (ETH)~\cite{ethereumhome}.

The novel semantics and programming model of smart contracts make
it challenging to ensure their correct behavior.
This makes them susceptible
to bugs or vulnerabilities that may be exploited by other accounts in the
Ethereum network. In fact, there has been a number of attacks on the
Ethereum main network that caused the loss of millions of ETH. The most
famous attack so far on Ethereum has been the attack on Decentralized
Autonomous Organisation (DAO), which was conducted by exploiting the
\emph{reentrancy} vulnerability. As a result of this attack, 3.5 million ETH
was stolen (about 50\,M USD at the time)~\cite{DASPTOP10a}.

Reentrancy involves repeated calls to the same function (or a set of
functions) before the first invocation is finished. Such nested invocations
can cause the smart contract to behave in unexpected ways, which can
be exploited by an attacker, usually to transfer funds away from the
victim contract. Reentrancy is known as one of the most dangerous
vulnerabilities in Ethereum smart contracts~\cite{atk-survey}.

Existing tools to detect reentrancy vulnerabilities use complex code
analysis and handcrafted rules to carefully analyze the control flow
and asset transfers in smart contracts. At a transaction level, such attacks are not explicitly observable, though. Our work attempts a completely new direction:

\begin{enumerate}
\item We monitor transactions at runtime at the level of the Ethereum
blockchain. A sample transaction trace that we gather data from for our machine learning model is presented in Figure \ref{fig1}. This monitoring does not require complex inspection of the smart
contracts themselves and makes it possible to deploy our technique
directly at the Ethereum block chain client, without any modification
of the smart contracts or the client involved.

\item We use machine learning on the monitored transaction metadata.
This avoids the need to design (possibly flawed) rules and also paves
the way towards recognizing new types of vulnerabilities in the future.
\end{enumerate}

Dynamit is designed to analyse transactions in smart contracts and report malicious ones.
Our technique, when used with a random-forest model, showed high accuracy
(96\,\%) on 105 transactions. We averaged our experimental
results over ten iterations of a setting that used ten-fold cross-validation for the training and test phases of all machine learning models per iteration.

The rest of this paper presents our work
in detail and is organized as follows: Section~\ref{sec:background}
explains smart contracts and reentrancy, and covers related work.
Section~\ref{sec:method} describes our approach. Our experiments and
their results are described in Sections~\ref{sec:experiments} and
\ref{sec:results}, respectively. Section~\ref{sec:threats} covers
threats to validity; Section~\ref{sec:conclusion} concludes and outlines
future work.

\begin{figure}[tb]
\begin{lstlisting}[language=Trace]
      blockHash: "0xb44...343c5",
      blockNumber: 33614,
      contractAddress: null,
      cumulativeGasUsed: 162534,
      from: "0x66a...1f3c7",
      gasUsed: 162534,
      logs: [],
      logsBloom: "0x0000...00000",
      status: "0x1",
      to: "0x89e...7ddc7",
      transactionHash: "0x3e4...039ad",
      transactionIndex: 0
\end{lstlisting}
\caption{A sample trace (transaction receipt) retrieved from the Ethereum blockchain}
\label{fig1}
\end{figure}

\section{Background}\label{sec:background}
\subsection{Ethereum}
Smart contracts embody a novel programming model that includes a global
shared state (managed in a decentralized way on a blockchain).
The
global state that stores everyone's assets is manipulated automatically
by the smart contracts, which are small programs that are expressed in
a specific format, such as
Ethereum bytecode~\cite{ethereumhome}.
This bytecode is usually compiled from a high-level
language, e.\,g., Solidity~\cite{dannen2017introducing}.
The code is executed by a virtual machine, instruction by instruction.
Each instruction also incurs a cost, measured in \emph{gas}, which the
invoker (user) of a smart contract has to pay. The VM
manages the effects of the instructions and their cost on everyone's
assets on the blockchain.
Potential vulnerabilities can arise at different levels in this
architecture; reentrancy is generally regarded as one of the most severe
ones~\cite{atk-survey,ashizawa2021eth2vec}.

\subsection{Reentrancy Vulnerability}
Contracts in Ethereum can send Ether to each other. Whenever a contract
receives a message without data that contains Ether and does not specify a
function, a default unnamed function, the
\textit{fallback} function, is invoked. 
When there is a transfer of funds
from contract $A$ to contract $B$, control will be handed over to
contract $B$~\cite{ethereumhome}. In the time
that $B$ has control, it can call back into any public function
of contract $A$, even the same function that issued the call to
$B$. This situation is called \emph{reentrancy}.

A simple example of reentrancy is illustrated in Figures~\ref{fig2} and~\ref{fig3}.
Contract \texttt{Vulnerable} donates to a target
contract (as specified by \texttt{to} parameter in \texttt{donate}). The intention is that a donation
occur only once, but this is not checked in \texttt{donate}.
An exploit is implemented by contract \texttt{Attacker}. Its \texttt{startAttack}
function issues a call to \texttt{donate} in
\texttt{Vulnerable}. After this call, \texttt{Vulnerable} transfers
plain Ether to \texttt{Attacker}. At this point, control is passed
to the fallback function of \texttt{Attacker}, which tries to call the
\texttt{donate} function again. The donations will continue until
\texttt{Vulnerable} runs out of gas or Ether, and because only
the last invocation is reverted upon failure, the attacker effectively
drains the victim of all its funds~\cite{atk-survey}.

In order to prevent reentrancy, one could use \textit{function modifiers}
in solidity~\cite{ethereumhome} to perform checks before giving the control to the fallback function of another contract. In case of our \texttt{Vulnerable}
contract, a simple check before sending the Ether prevents
the attacker from exploiting reentrancy. The safe version, \texttt{NotVulnerable},
checks and updates its state before sending Ether to the interacting contract
(see Figure~\ref{fig4}).

\begin{figure}[tb]

\begin{lstlisting}[language=Solidity]
   contract Vulnerable {
      function donate(address to_) public payable 
      {
         require(to_.call.value(1 ether)());
      }
   }
\end{lstlisting}
\caption{A sample contract vulnerable to attacks on reentrancy}
\label{fig2}
\end{figure}

\begin{figure}[tb]
\begin{lstlisting}[language=Solidity]
   contract Attacker {
      Vulnerable public vul_contract;
      function startAttack(address _addr) public 
      {
         vulContract = Vulnerable(_addr);
         vulContract.donate(address(this));
      }
      function() public payable 
      {
         vulContract.donate(address(this));
      }
   }
\end{lstlisting}
\caption{A sample contract that exploits the vulnerable contract}
\label{fig3}
\end{figure}

\subsection{Related Work}

Program analysis techniques to detect potential vulnerabilities
can be divided into \emph{static analysis}, which analyzes
the structure of code without running it, and \emph{dynamic analysis},
which analyzes the runtime behavior of an executing program.
The advantage of static analysis is that it does not require a test case
to reveal a flaw; conversely, it has the disadvantage that the
analysis may be overly strict and reveal \emph{spurious} problems that
are not actually exploitable flaws during program execution. Dynamic
analysis, on the other hand, always produces actual executions (and thus
a witness of a real problem), but may be unsuccessful at finding the right
inputs to make this happen.
Combinations
of these techniques also exist, typically in the form of static analysis to identify
parts of the program that might need closer inspection at runtime.

\begin{figure}[tb]
\begin{lstlisting}[language=Solidity]
   contract NotVulnerable {
      mapping (address => bool) public donated;
      function donate(address to_) public payable 
      {
         if (donated[to_] != true) {
            donated[to_] = true;
            require(to_.call.value(1 ether)());
         }
      }
   }
\end{lstlisting}
\caption{A sample contract with the vulnerability 
fixed}
\label{fig4}
\end{figure}

\emph{Static analysis} tools for smart contracts include tools like
Securify~\cite{tsankov2018securify}, SmartCheck~\cite{tikhomirov2018smartcheck},
and Slither~\cite{feist2019slither}. These tools check code against
problematic patterns that may constitute violations of coding guidelines
or even potential vulnerabilities.
Symbolic execution is a static analysis technique that uses path conditions
(conditions about the feasibility of certain execution paths) to reason
about inputs that may reach a potentially unsafe state in a program.
Oyente~\cite{luu2016making} was the first tool to apply symbolic execution
to smart contracts. Other tools, such as TeEther~\cite{krupp2018teether},
MAIAN~\cite{nikolic2018finding}, and Zeus~\cite{kalra2018zeus},
followed and took different approaches
at finding harmful inputs or detecting problems in the contracts.

The above-mentioned static analysis tools use rules designed by
experts to detect problems. Recent works have adapted the use
of machine learning to static analysis~\cite{shin2015recognizing}
and extended this idea to the domain of smart contracts~\cite{ashizawa2021eth2vec}.

\emph{Dynamic analysis} for smart contracts focuses on finding inputs that reach a program state
exhibiting a problematic execution pattern (such as reentrancy),
implemented in tools like Echidna~\cite{grieco2020echidna},
ContractFuzzer~\cite{jiang2018contractfuzzer}, and ReGuard~\cite{liu2018reguard}.
The accuracy of these tools depends on the quality of the hand-coded
pattern recognizer.
Recent work uses an oracle that tracks the \emph{balance} of each
smart contract instance to detect fundamental misuse, thus eliminating
the need for specific program patterns to detect a vulnerability~\cite{wangOracleSupportedDynamicExploit2020}.

Our work leverages machine learning to detect problematic
execution patterns, focusing on reentrancy in smart contracts. It is,
to our knowledge, the first work to analyze dynamic execution patterns
in smart contracts through machine learning. In the area of malware detection, a metadata-focused approach has also
been used successfully~\cite{afzal2020using} by looking at the frequency
and size of packets sent over encrypted connections to classify the
behavior of an application. 

\section{Method}\label{sec:method}
The Dynamit framework detects reentrancy vulnerabilities in deployed smart contracts without needing their source code. Dynamit considers only the dynamic behavior of the smart contract; that behavior is extracted from metadata describing the transactions between the contracts. This monitoring is based on the existing application programming interface (API) of the unmodified Ethereum blockchain client.

\begin{figure}[tb]
\centerline{\includegraphics[scale=0.85]{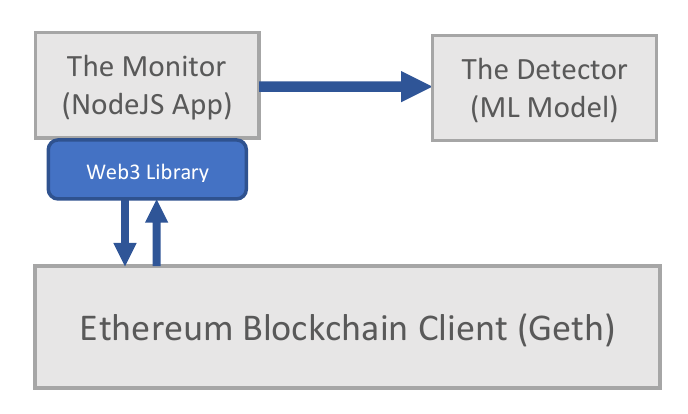}}
\caption{Dynamit framework system diagram}
\label{fig5}
\end{figure}

Dynamit consists of two parts (see Figure~\ref{fig5}):
\begin{enumerate}
  \item The \emph{Monitor,} which observes transactions in the blockchain.
  \item The \emph{Detector,} which classifies behavior as benign or malicious.
\end{enumerate}
The detector can be configured against various classifiers, which are first trained on a training set before our tool is put to use to detect malicious transactions in production.

\subsection{Monitor}
The monitor connects to the Ethereum blockchain client to gather information about desired transactions. It uses the latest version of Web3js \cite{Web3JsEthereum}, which is the official Ethereum Javascript API to connect to and probe the Ethereum network.

The monitoring obtains the data as follows:

\begin{itemize}
  \item \textbf{Subscription to the events emitted by the Ethereum client}. These events are emitted when a transaction related to an account is issued \cite{Web3JsEthereum}. In our work, we use \textit{pendingTransactions} to get any new transactions related to our accounts under observation.
  \item \textbf{Probing the blockchain at specific intervals until the desired information is retrieved}. This is suitable for getting information about an already mined transaction or getting the state of a contract after an event. 
\end{itemize}

\subsection{Detector}
The detector is the part of the system that distinguishes the harmful transactions from the benign ones. It consists of a part that processes and cleans that data received by the monitor, and a machine learning model that is trained as the monitor feeds in the data. 

\subsubsection{Extracted Features}
The extracted features and the mechanism used to monitor them are presented in Table~\ref{tab1}.

\begin{table}[tb]
\caption{Features gathered by monitor and used by our ML model in detector}
\begin{center}
{\small
\begin{tabular}{ll}
\toprule
\textbf{Feature} & \textbf{Monitoring Mechanism} \\
\midrule
Gas usage of transaction & Event subscription\\

Contract 1 balance difference & Probing\\

Contract 2 balance difference & Probing \\

Average call stack depth & Probing\\
\bottomrule
\end{tabular}
\label{tab1}
}
\end{center}
\end{table}

The \textit{contract balance difference} feature is the difference of the balance of a contract before and after the transaction has taken place. In fact, the feature \textit{contract balance difference} may easily be replaced by any other asset that is being transferred by the contracts to match the specific use case.

The \textit{average call stack depth} is the only feature being directly retrieved from the transaction trace. Calling a regular function in a contract will not dramatically change the value of this feature. However, recursive external calls will drastically change this value. This is often the case for the reentrancy vulnerability, where a particular function in victim is recursively called until the attacker contract stops. Intuitively, this feature should have a positive correlation with the transaction being harmful. However, an attacker can easily avoid being detected by limiting the number of recursions. Therefore, we decided to put effort to randomly decrease the average call stack depth for harmful transactions to make it harder for the detector to distinguish them, and to decrease the bias of the model. 

As mentioned earlier, as contract code executes on the blockchain, it consumes gas. Gas usage depends on the specific operations that a contract carries out within a transaction. Since a successful attack on a vulnerable contract may exhibit a specific execution pattern, we use this gas usage as a summary representation of the execution.

\subsubsection{Classifier}
To find the best model, we trained and tested the following models in our detector:

\begin{itemize}
  \item \textbf{Random Forest}: A random-forest (RF) classifier using bagging with 100 decision trees.
  \item \textbf{Naive Bayes (NB)}
  \item \textbf{Logistic Regression}
  \item \textbf{K-Nearest Neighbours}: A K-NN with 5 neighbors.
  \item \textbf{Support Vector Machine (SVM)}: Both linear and polynomial kernels were used. The model with linear kernel out-performed the other one. 
\end{itemize}
All of our models were built using Scikit-learn library~\cite{pedregosa2011scikit}. The Random Forest model is composed of 100 trees of type DecisionTreeClassifier in Scikit-learn.

\subsection{Usage of Dynamit}
Let us assume developers of an application deploy it as a smart contract on top of Ethereum. Dynamit can be used by these developers to safeguard their smart contract. They install Dynamit on their own machine, and configure it to connect to Ethereum network to monitor their deployed contract.
As the transactions are issued to the monitored smart contract, Dynamit collects and processes their metadata. The previously trained machine learning model then classifies transactions as benign or harmful; the latter can be used as feedback to the developer or as part of a security information and event management (SIEM) system that may report users to an administrator or block a vulnerable contract from being used further. 

\section{Experiments}\label{sec:experiments}

\begin{table}[tb]
    \centering
    {\small
    \begin{tabular}{l l}
    \toprule
    \textbf{Service contract} & \textbf{User contract}\\
    \midrule
        13 robust contracts & 11 benign contracts \\
        12 vulnerable contracts & \hphantom{0}9 malicious contracts\\
         \bottomrule
    \end{tabular}
    }
    \caption{Set of contracts used in the experiments}
    \label{tab:contracts}
\end{table}

We chose 25 open-source contracts for our experiments that implement a certain functionality that we denote as \emph{service contracts} here. These contracts were originally used in \cite{li_vuldeepecker}. Their source code is available on Etherscan.\footnote{\url{https://etherscan.io/}} We wrote 20 \emph{user contracts} that access and utilize that functionality (see Table~\ref{tab:contracts}). 

A service contract may be robust (not exploitable) or contain a vulnerability; likewise, a user contract may be benign or malicious. Only a combination of a vulnerable service contract with a malicious user may actually reveal the vulnerability in the service contract.

For the experiment, we monitored a total of 105 transactions generated from these contracts, with 53 benign and 52 harmful transactions. All of these transactions have been labelled manually before starting the experiment, so they could be used for both training and testing a supervised model.
We feed labelled transaction data to our classifier (offline) for the training phase; in production, online (unlabelled) data can be used.

From the 105 transactions, 25 transactions were taken from the 25 open-source service contracts, which we complemented with 20 variants of user contracts.
The remaining 80 transactions are generated using two pairs of contract templates (four contracts) that generate both harmful and benign transactions randomly. Contract \texttt{Vulnerable2} is one such variant of a service contract, which donates a random amount to the user (see Figure~\ref{fig6}). To generate these random transactions, both the service and the user contracts (see Table \ref{tab:contracts}) fuzz their behavior to represent different behaviors of real-world scenarios. Another reason for having random behavior (fuzzing) in both service and user contracts is that there may be a complex internal computation that has a certain call stack depth or gas usage. This potentially can make an attack harder to detect. We would like to have such behavior included in our data to have a less biased classifier in detector.
Therefore, these transactions are generated in a way to prevent overfitting in the model. For example,
we \emph{fuzz} the gas usage by injecting a random loop with 50\,\% probability in the vulnerable contract template (see lines 12--18 in Figure~\ref{fig6}). Each use of the counter expends extra gas.
Likewise, we randomize the amount donated to the user and the number of times an
attacker actually exploits reentrancy, to make it harder to recognize the attacks.


Since each interaction between a service and its user is either benign or harmful, the following outcomes can occur:


\begin{itemize}
  \item The user contract successfully exploits the reentrancy vulnerability: \textit{harmful transaction.}
  \item The user contract tries to exploit a reentrancy vulnerability (which may or may not exist in the service contract) but is unsuccessful. This will lead to one of the following situations:
  \begin{itemize}
     \item The transaction and accordingly its effects on the target contract state are reverted by the Ethereum runtime environment. Such failed (reverted) transactions are not made visible through the monitoring API in Ethereum and therefore not taken into account by our analysis.
     \item The transaction is not reverted, and takes the intended original effect: \textit{benign transaction.}
   \end{itemize}
  \item The peer contract does not try to exploit reentrancy at all: \textit{benign transaction.}
\end{itemize}

As mentioned earlier, after data is collected by the monitor, it will be fed in to the detector for classification. We trained and tested the models in detector using above-mentioned data. For all of our models, we used stratified 10-fold cross-validated training and test sets to get consistent and reliable results. For each number in the plots, the whole experiment (including the cross validation) has been performed 10 times, and the average performance was taken. The numbers of neighbors in K-NN model and number of trees in our RF model are chosen based on empirical observations to maximize the performance of the model.

\begin{figure}[tb]
\begin{lstlisting}[language=Solidity]
  contract Vulnerable2 {
    uint public gasFuzzingCounter = 0;
    uint public c = 0;
    uint public d_binary = 0;
    uint public amnt;
    function random(uint num) private view returns (uint8) {
       return uint8(uint256(keccak256(block.timestamp, block.difficulty))%num);
    }
    constructor() public payable {
    }
    function donate(address to_) public payable {
       d_binary = random_binary();
       c = random(10);
       if (d_binary == 1) {
          for (uint i = 0; i < c; i++) {
             gasFuzzingCounter++;
          }
       }
       amnt = random(1000) * 500000000000000;
       require(to_.call.value(amount)());
    } 
  }
\end{lstlisting}
\caption{Source code for the smart contract used to generate random harmful transactions for the experiment}
\label{fig6}
\end{figure}

\section{Results}\label{sec:results}
\begin{figure}[tb]
\centerline{\includegraphics[width=9.4cm]{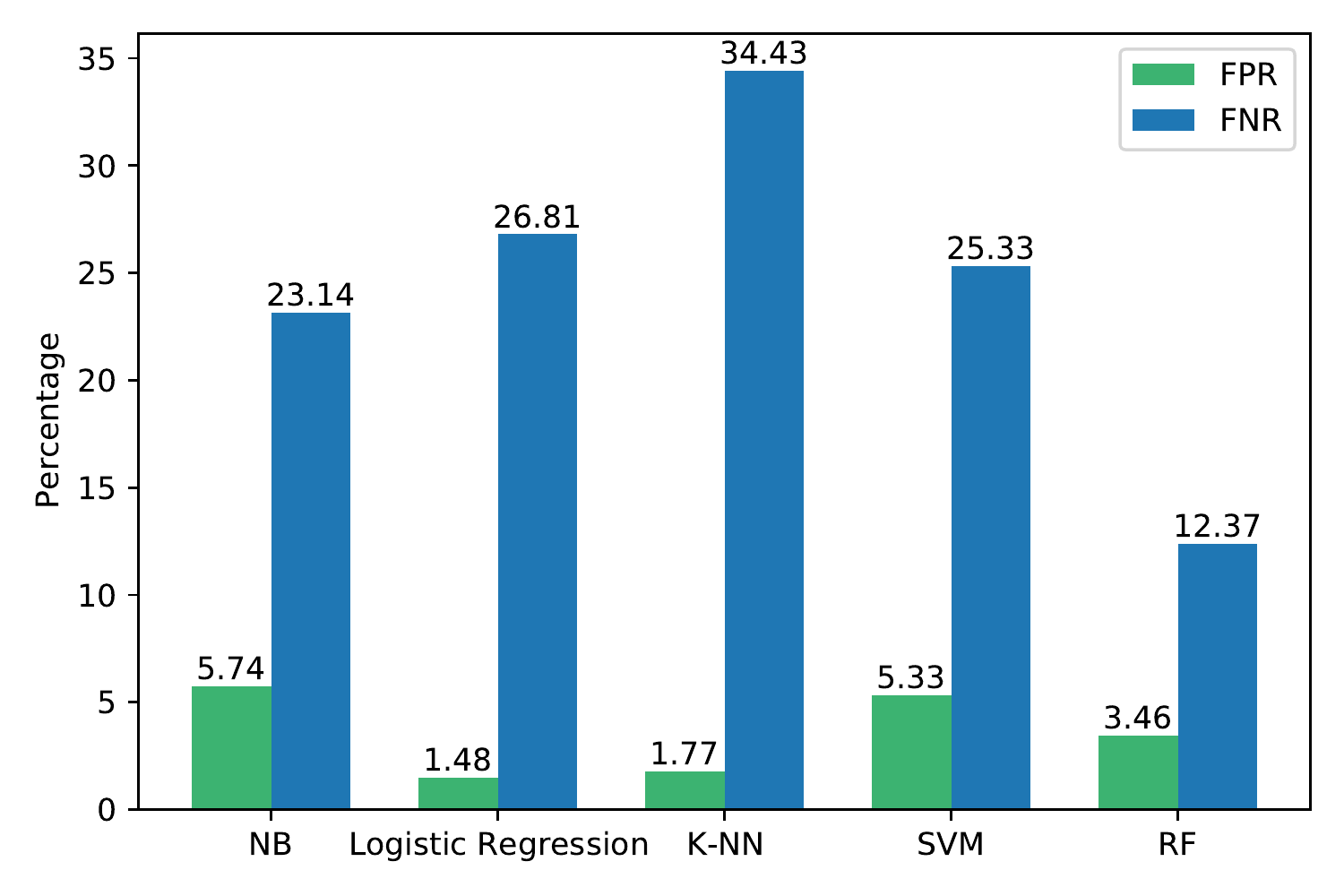}}
\caption{Average false positive rate and false negative rate for detecting vulnerable transactions with different classification models.}
\label{fig7}
\end{figure}

\begin{figure}[tb]
\centerline{\includegraphics[width=9.4cm]{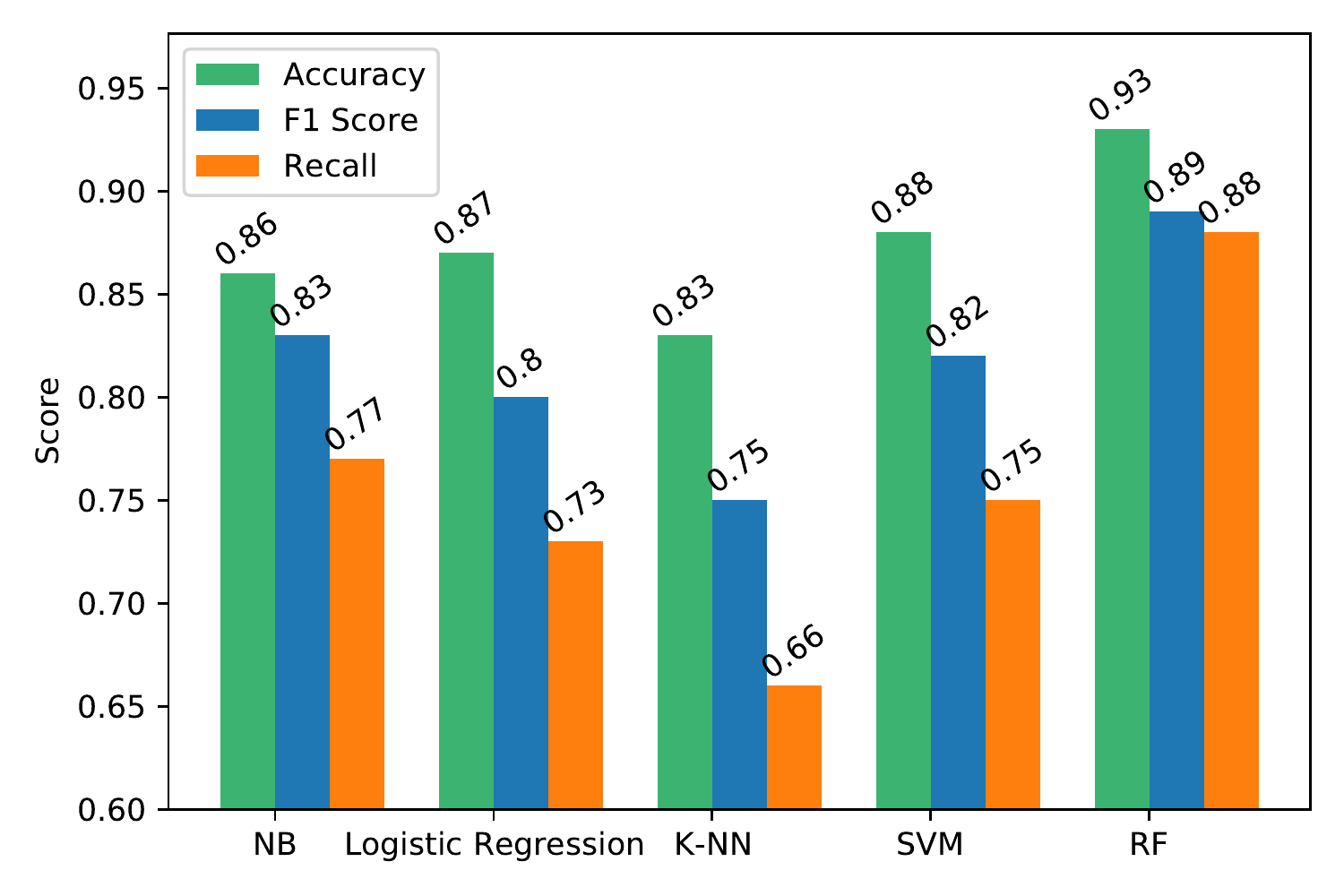}}
\caption{Average Accuracy, F1 Score, and Recall for detecting vulnerable transactions with different classification models.}
\label{fig8}
\end{figure}

We train and test five different types of classifiers and compare them based on the average false positive rate (FPR) and false negative rate (FNR) as well as accuracy, F1 score, and recall (see Figures~\ref{fig7} and~\ref{fig8}). The FPR varies between 1.48\,\% (logistic regression) and 5.74\,\%(Naive Bayes), while the FNR is the lowest for the random forest (RF) model at 12.37\,\%.

The RF classifier achieves the highest accuracy (93\,\%).
Most of the inaccuracy of the models can be attributed to the FNR. In other words, the detector is labelling a considerable number of harmful transactions as benign (even using RF). Conversely, the low FPR makes Dynamit useful
as a monitoring tool in scenarios where the cost of false positives are rather high, such as in testing or when suspending problematic contracts in production for manual review.

\begin{figure}[tb]
\centerline{\includegraphics[width=9.2cm]{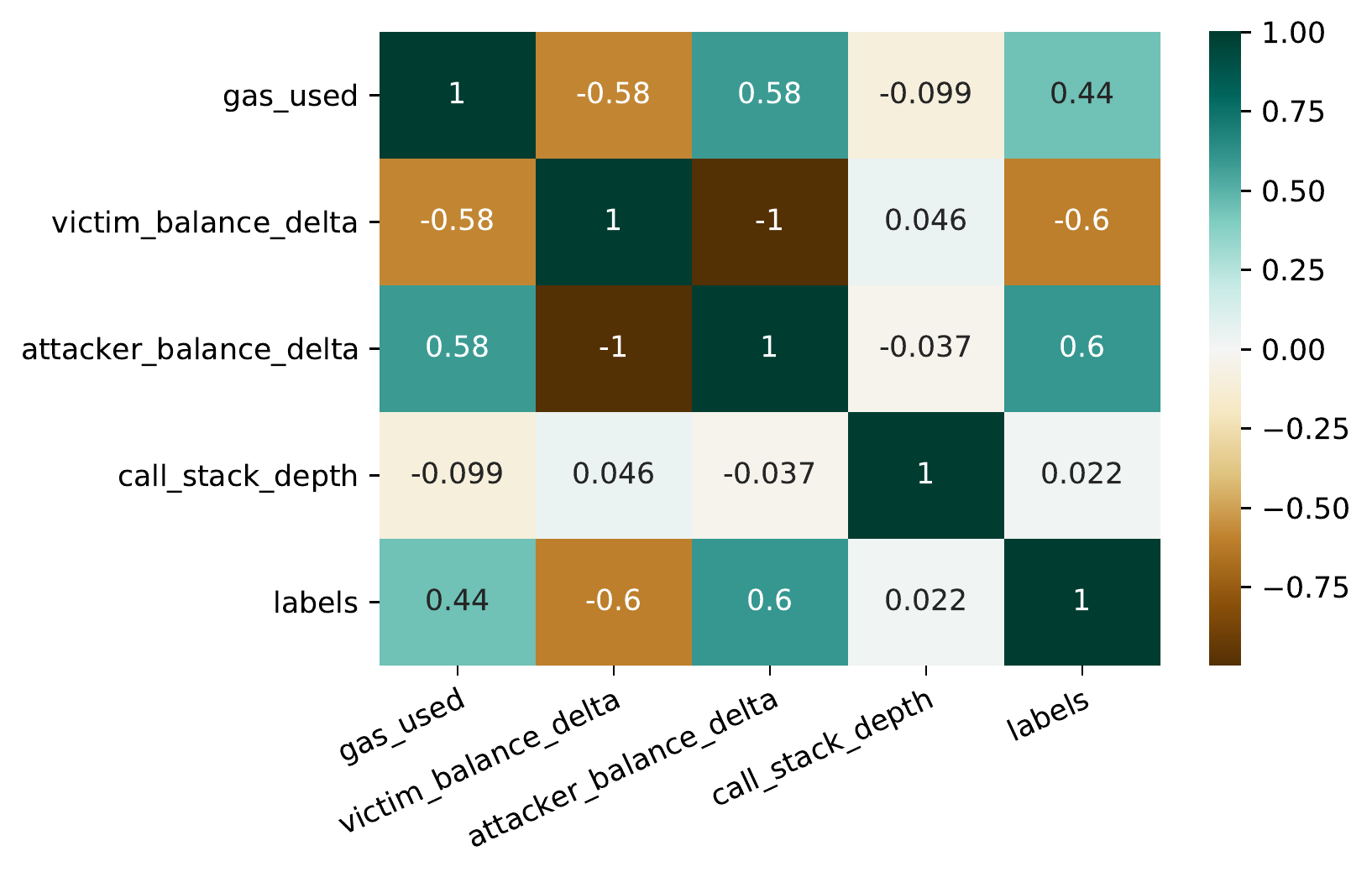}}
\caption{Feature correlation heatmap for the detectors with label  1 for harmful transactions and 0 for benign ones.}
\label{fig9}
\end{figure}

\begin{figure}[tb]
\centerline{\includegraphics[width=9.4cm]{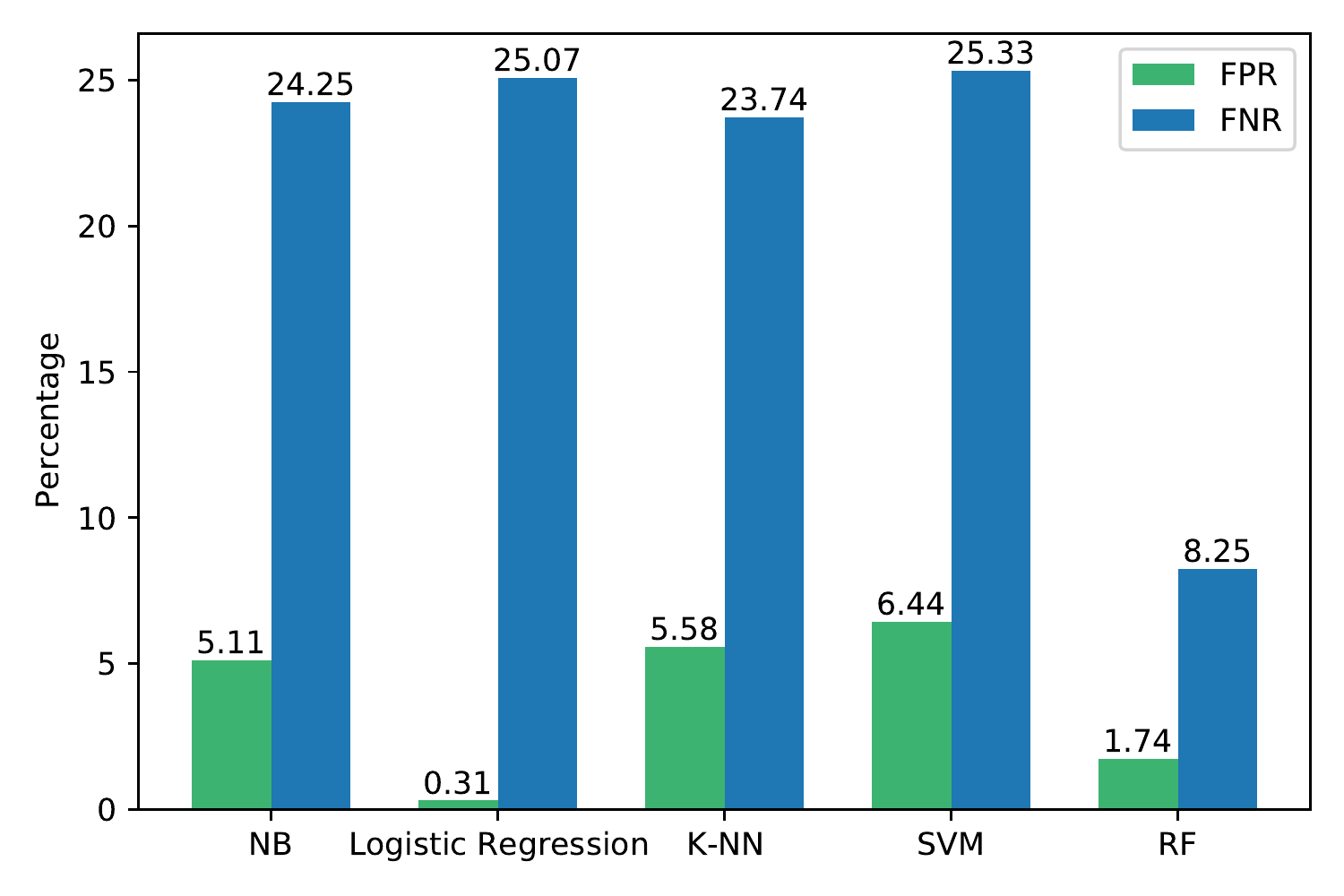}}
\caption{Average false positive rate and false negative rate for detecting vulnerable transactions with different classification models without the \textit{average call stack depth} feature.}
\label{fig10}
\end{figure}

\begin{figure}[htbp]
\centerline{\includegraphics[width=9.4cm]{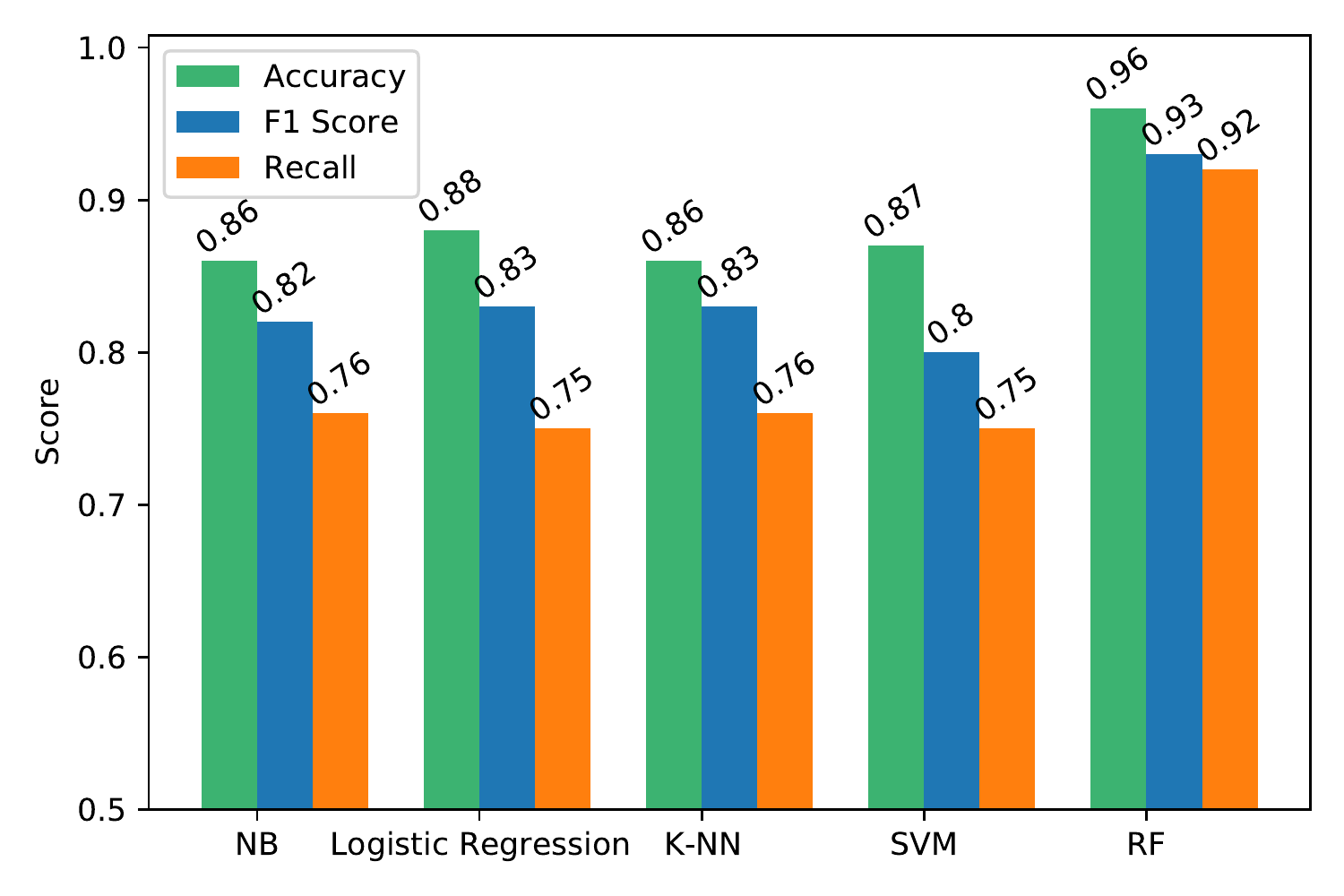}}
\caption{Average Accuracy, F1 Score, and Recall for detecting vulnerable transactions with different classification models without the \textit{average call stack depth} feature.}
\label{fig11}
\end{figure}

As mentioned earlier, the contract sets we used for random transaction generation try to disguise their behavior. We took this measure
to build a realistic model and decrease bias. 
As a result of this, the correlation of the \textit{average call stack depth} and the label of the transaction is very low (see Figure \ref{fig9}). Hence, we decided to also build the same models without the \textit{average call stack depth} feature. The results of this version of the models are shown in Figures~\ref{fig10} and~\ref{fig11}. The overall behavior of all models is consistent with results in Figures~\ref{fig7} and~\ref{fig8}. However, there are a few interesting changes. While RF is still the most accurate model and even more accurate than before, the relative reduction in the FPR of RF is higher than the one in the FNR. The highest average accuracy in this experiment belongs to RF (96\,\%).

\section{Threats to Validity}\label{sec:threats}

We use a total number of 49 smart contracts (25 serivce, 20 user, and 4 random transaction generation contracts) in our experiments. In an effort to collect more realistic data, the harmful transactions issued by our own contracts are randomized to disguise their malicious nature. As mentioned in results section, this has rendered our otherwise important \textit{average call stack depth} feature useless by making the \textit{average call stack depth} of a harmful contract seem like a benign one, and vice versa. From a different point of view, this also shows the possibility of tricking a dynamic detector if it only uses checks on certain variables (such as the balance) to detect a vulnerability. Our results suggest that a combination of our machine learning-based detector and an oracle-supported dynamic vulnerability detection~\cite{wangOracleSupportedDynamicExploit2020} may decrease the number of false negatives.

Another consideration is the amount of randomness that our randomly generated transactions have. In case there is not enough randomness, our machine learning model in detector will exhibit high bias, rendering it useless for catching more complex attacks. Our random transaction generator uses the block's timestamp and difficulty to generate random numbers. Since we have used a private deployment of Ethereum blockchain, the mentioned variables were controllable by increasing the mining frequency and issuing transaction generation commands each 30 seconds. Using this method, we verified that the random transaction generation system has enough randomness.

\section{Conclusion and Future Work}\label{sec:conclusion}
 In this work, we present Dynamit, a dynamic vulnerability detection framework for Ethereum smart contracts. Dynamit detects vulnerable smart contracts by classifying harmful transactions in a blockchain using machine learning on transactional metadata. We achieve 96\,\% accuracy on a data set of 105 transactions.

To further develop Dynamit, we will
investigate automatic test-case generation tools such as Vultron
\cite{vultron}. Such tools can generate labeled transactions and create benign and malicious user contracts to reproduce them. Another direction for future work
is to find more features to make the detection more accurate. An
example would be to observe bookkeeping variables inside the contracts,
and the way they change, as additional indicators of a smart contract
being exploited.
Finally, we will consider analyzing sequences of multiple transactions and applying other types of machine learning to the data, to increase the capabilities of our detector and to analyze other types of vulnerabilities as well.

\bibliographystyle{named}
\bibliography{paper}

\end{document}